\documentclass[reprint, superscriptaddress, amsmath, amssymb, aps, nofootinbib, floatfix]{revtex4-2}

\usepackage{graphicx}
\usepackage[colorlinks=true,urlcolor=blue,linkcolor=blue,citecolor=blue,hypertexnames]{hyperref}
\usepackage[nameinlink]{cleveref}
\usepackage{tikz}

\newcommand{\order}[1]{{\cal O}(#1)}
\newcommand{\pderiv}[2]{\frac{\partial #1}{\partial #2}}
\newcommand{\dderiv}[2]{\frac{\dd #1}{\dd #2}}

\newcommand{\fderiv}[2]{\frac{\delta #1}{\delta #2}}
\newcommand{\dd}{\text{d}}

\crefname{section}{Sec.\!}{Secs.\!}
\crefname{figure}{Fig.\!}{Figs.\!}
\crefname{equation}{}{}
\crefname{table}{Tab.\!}{Tabs.\!}
\crefname{appendix}{App.\!}{Apps.\!}

\begin{document}

\title{Renormalization group flows from the Hessian geometry of quantum effective actions}

\author{Yannick Kluth}
\email{yannick.kluth@manchester.ac.uk}
\affiliation{Department of Physics and Astronomy, University of Manchester, Manchester M13 9PL, United Kingdom}
\affiliation{Perimeter Institute for Theoretical Physics, Waterloo, Ontario N2L 2Y5, Canada}
\author{Peter Millington}
\email{peter.millington@manchester.ac.uk}
\affiliation{Department of Physics and Astronomy, University of Manchester, Manchester M13 9PL, United Kingdom}
\author{Paul Saffin}
\email{paul.saffin@nottingham.ac.uk}
\affiliation{School of Physics and Astronomy, University of Nottingham, Nottingham NG7 2RD, United Kingdom}

\date{28 November, 2023}

\begin{abstract}
We explore a geometric perspective on quantum field theory by considering the configuration space, where all field configurations reside. Employing $n$-particle irreducible effective actions constructed via Legendre transforms of the Schwinger functional, this configuration space can be associated with a Hessian manifold. This allows for various properties and uses of the $n$-particle irreducible effective actions to be re-cast in geometrical terms. In particular, interpreting the two-point source as a regulator, this approach can be readily connected to the functional renormalization group. Renormalization group flows are then understood in terms of geodesics on this Hessian manifold.
\end{abstract}
\maketitle

\tableofcontents

\section{Introduction}

Understanding physical concepts within a geometric framework has demonstrated its utility across various domains of physics. The most prominent example might be Einstein's general theory of relativity, wherein gravity is described in terms of the geometry of spacetime. However, geometrical ideas have also been applied to other areas of physics, including quantum field theory (QFT).

QFT operates within an infinite-dimensional configuration space that encompasses all conceivable field configurations. The pursuit of non-trivial geometric structures in this space has been a longstanding endeavor, exemplified by the Vilkovisky--DeWitt connection~\cite{Vilkovisky:1984st, DeWitt:1985sg}, which allows the definition of a covariant derivative within this configuration space. Extensions of this framework then allow to write the path integral in a form that is manifestly invariant under both spacetime and field-space diffeomorphisms~\cite{Finn:2019aip, Finn:2020nvn, Gattus:2023gep}. More recently, the introduction of a covariant structure has also been used to show the invariance of field-theoretic observables under field redefinitions~\cite{Cohen:2022uuw}.

In statistical mechanics, geometric approaches have a long history, going back to the work of Ruppeiner~\cite{ruppeiner1979thermodynamics}, building on that of Weinhold~\cite{10.1063/1.431689, 10.1063/1.431635, 10.1063/1.431636, 10.1063/1.431637}, on the representation of thermodynamic systems by Riemannian manifolds. Therein, the metric can be defined in terms of the Hessian of the entropy with respect to the extensive state variables~\cite{Diosi:1983wa}, viz., the two-point correlation functions. Such considerations then lead naturally to the application of information theoretic techniques to the renormalization group (RG) (for a comprehensive discussion of information-geometrical approaches to QFT, see~\cite{Floerchinger:2023ekw}).

In this context, the importance of considering the geometry of the theory space is perhaps epitomized by Zomolodchikov's $C$-theorem~\cite{Zamolodchikov:1986gt}, and the study of information loss along RG flows in the space of couplings has attracted much attention~\cite{Dolan:1997cx, Brody:1997gn, Dolan:1999gk, Apenko:2009kq, Beny:2012qh, Beny:2014sna, Gaite:2000zr, Koenigstein:2021rxj}. Here, the Fisher information metric plays a key role~\cite{OConnor:1993dvz}, and one can make connection between the relative entropy and the one-particle irreducible (1PI) effective action~\cite{Gaite:1995yg}.  The geometrical interpretation of renormalization is then such that the couplings (viz.\ one-point sources) are understood as coordinates and the associated composite operators as tangent vectors on the manifold of the theory space~\cite{Lassig:1989tc}; the RG equations describe the transport generated by the beta functions~\cite{doi:10.1142/S0217751X94000571}. 

In this work, we take a different approach:\ Rather than the space of couplings, we use a geometrical framework on the configuration space of QFT --- the space of all field configurations --- which includes $n$-point sources, i.e., sources that depend on $n$ spacetime coordinates. A crucial feature is the emergence of an apparent dual structure present in any QFT. This duality emerges through Legendre transforms that link the Schwinger functional to various quantum effective actions. Sources in the Schwinger functional are mapped to expectation values, both of which fully characterize the configuration space. Geometrically, these structures can be elegantly implemented through information-geometrical manifolds~\cite{amari1980theory,amari1982differential,chentsov1982statiscal,nielsen2020elementary, Floerchinger:2023ekw, Floerchinger:2023qpw}. The configuration space of QFT can then be conceived as a Hessian manifold~\cite{koszul1961domaines,shima1976certain,cheng1982real,shimabook}, where the metric is given by the Hessian of a potential. This potential aligns with the Schwinger functional, and the metric with the two-point functions of all composite operators to which the sources couple. Furthermore, the inverse two-point functions are obtained from the Hessian of a dual potential --- the quantum effective action.

We use these geometrical insights to examine the $n$-particle irreducible ($n$PI) and related quantum effective actions. These involve sources not only for the bare quantum field in the Schwinger functional but also for up to the $n$-point function. Going beyond local sources in this way, cf.~\cite{Lassig:1989tc}, this naturally relates to the functional renormalization group (FRG), where the source for the two-point function acts as a regulator~\cite{Polchinski:1983gv, Wetterich:1992yh, Morris:1993qb, Ellwanger:1993mw, Rosten:2010vm, Dupuis:2020fhh}. By incorporating this structure within the Hessian manifold, we gain a novel interpretation of the resulting RG flow from a geometric standpoint. In fact, RG trajectories can be shown to correspond to certain geodesics that arise naturally on the Hessian manifold.

This article is structured as follows. In Sec.~\ref{sec:nPIEAs}, we set up our notation for the $n$PI effective actions, before introducing the associated Hessian manifolds and their structures in Sec.~\ref{sec:hessman}. We subsequently employ the Hessian geometry to construct RG flows in Sec.~\ref{sec:RG}, making connections with other formulations of the FRG.  Section~\ref{sec:conc} provides our concluding remarks. Further technical details are provided in the appendices. Appendix~\ref{sec:HesseKoszul} reviews the Hesse--Koszul~\cite{MIRGHAFOURI201754, ARXIV.2001.02940} flow of the metric of the Hessian manifold, which bears some similarity to the Wetterich equation~\cite{Wetterich:1992yh, Morris:1993qb, Ellwanger:1993mw}. The closure of the 2PI RG is described in App.~\ref{sec:closure}, and subtleties related to the non-commutativity of various derivatives are described in App.~\ref{sec:derivs}

%%%%%%%%%%%%%%%%%%%%

\section{The \texorpdfstring{$n$}{n}PI Effective Actions}
\label{sec:nPIEAs}

In this section, we define the family of $n$PI effective actions and introduce our notation. We start with the definition of the partition function and consider a QFT with a single scalar field $\Phi$, but all expressions below can be readily generalized by the addition of a contracted multi-index that accounts for other degrees of freedom or quantum numbers. 

To define the partition function suitable for the $n$PI effective action, we start with $n$-point sources
\begin{equation}
J^{(n)}=J_{n}(x_1,x_2,\dots,x_n)\,.
\end{equation}
Each of these sources couples in the path integral to composite operators of the form
\begin{equation}
    \Phi^{(n)}=\Phi(x_1)\Phi(x_2)\cdots \Phi(x_n)\;.
\end{equation}
The partition function can then be defined by
\begin{equation}
    \mathcal{Z} [\{ J \}_n] = \int \mathcal{D} \Phi \, \exp \left\{ - S [\Phi] + \sum_{i = 1}^n J^{(i)} \cdot \Phi^{(i)} \right\} \, ,
\end{equation}
where the ``$\cdot$'' represents integration over spacetime variables included in the sources and field variables, and we use $\hbar = 1$. The notation ``$\{ J \}_n$'' indicates that we consider the generating functional with up to $n$-point sources. Starting from the partition function, the Schwinger functional is given by
\begin{equation}
    \mathcal{W}[\{J\}_n]=\ln\mathcal{Z}[\{J\}_n]\;.
\end{equation}

The effective action can be obtained from the Schwinger functional by a Legendre transform. Here, we allow for generalized effective actions that are obtained by Legendre transforms with respect to not necessarily all sources $J^{(n)}$ of the Schwinger functional, but only a subset of them. The resulting family of effective actions can be denoted by
\begin{equation}
    \Gamma[\{\Delta^{(i)}\:|
\:i\in I\};\{J^{(j)}\:|\: j\notin I\}]=\Gamma[\emptyset;\{J\}_n]+\sum_{i\:\in\: I}J^{(i)} \cdot \Delta^{(i)}\;,
\label{eqn:genEAA}
\end{equation}
wherein we identify
\begin{equation}
\Gamma[\emptyset;\{J\}_n]=-\mathcal{W}[\{J\}_n] \, .
\end{equation}
In \cref{eqn:genEAA}, the Legendre transform is with respect to all sources $J^{(i)}$ of the Schwinger functional that are included in the set $I\subset\{1,2,\dots,n\}$. Sources $J^{(i)}$ that are not included remain as arguments of the effective action. We denote this in \cref{eqn:genEAA} by a semicolon that separates variables that were included in the Legendre transform from variables that were not. Excluding $I = \emptyset$, we find $2^n-1$ different effective actions depending on which sources are in included in the Legendre transform. Any $i$-point source $J^{(i)}$ that is included in the Legendre transform is eliminated in favour of the disconnected $i$-point function $\Delta^{(i)}$ in the effective action, with
\begin{equation}
    \Delta^{(i)}=\Delta(x_1,x_2,\dots,x_i) \equiv \langle \Phi (x_1) \, \Phi (x_2) \, \dots \Phi (x_i) \rangle\;,
\end{equation}
such that $\Delta^{(1)}=\langle\Phi(x_1)\rangle$ is the one-point function. In this notation, the $n$PI effective action is
\begin{equation}
    \Gamma[\{\Delta^{(i)}\}_n;\emptyset]=\Gamma[\emptyset;\{J\}_n]+\sum_{i\,=\,1}^nJ^{(i)} \cdot \Delta^{(i)}\;.
\end{equation}

Let us consider the case $n = 2$ more specifically.
Starting with the partition function, we introduce sources $J^{(1)} = J$ and $J^{(2)} = K$ for the one- and two-point function, respectively, writing
\begin{equation}
    \mathcal{Z} [J, K] = \int \mathcal{D} \Phi^a e^{-S[\Phi^a] + J_a \Phi^a + \Phi^a K_{a b} \Phi^b} \,.
    \label{eqn:Zdef}
\end{equation}
Herein, lower-case Latin characters are  DeWitt indices, which, in general, include both a spacetime coordinate and discrete indices. The continuum components of repeated DeWitt indices are integrated over spacetime, and discrete components are summed over. In the following, we also introduce the convention that Greek letters correspond to pairs of lower-case Latin indices, i.e.,
\begin{equation}
    K_\alpha = K_{a b} \, .
\end{equation} The resulting Schwinger functional from \cref{eqn:Zdef} is given by
\begin{equation}
    \mathcal{W} [J, K] = \ln \mathcal{Z} [J, K] \, .
    \label{eqn:Wdef}
\end{equation}

The inclusion of a source for the two-point function is particularly useful for the studies of the FRG on which this work will focus later. Namely, the $K_\alpha$ can be associated with a regulator term. Varying the size of this regulator, we can then obtain a flow for the action at different RG scales $k$ induced by $K_\alpha$. Slices of constant $K_\alpha$ then correspond to configurations of equal RG scale. At any given value of $K_\alpha$, the coordinates encoded by $J_a$ are related to connected $n$-point functions (denoted by a subscript c) evaluated at a given RG time by 
\begin{equation}
    \langle \phi^{a_1} \dots \phi^{a_n} \rangle_{\text{c}} \equiv \fderiv{}{J_{a_1}} \dots \fderiv{}{J_{a_n}} \mathcal{W} [J, K] \, .
    \label{eqn:npointdef1}
\end{equation}

The family of $n=2$ effective actions arising from \cref{eqn:genEAA} contains four functionals:
\begin{subequations}
\label{eq:fourpotentials}
\begin{align}
    \Gamma[\emptyset;J,K]&=-\mathcal{W}[J,K]\;,\\
    \Gamma[\phi;K]&=-\mathcal{W}[J,K]+J_a\phi^a\;,\\
    \Gamma[\Delta;J]&=-\mathcal{W}[J,K]+K_\alpha\Delta^\alpha\;,\\
    \Gamma[\phi,\Delta;\emptyset]&=-\mathcal{W}[J,K]+J_a\phi^a+K_\alpha\Delta^\alpha\;,
\end{align}
\end{subequations}
where the disconnected $n$-point functions are denoted by $\Delta^{(1)} = \phi$, and $\Delta^{(2)} = \Delta$. The Schwinger functional is given by $\Gamma[\emptyset;J,K]$, and $\Gamma[\phi,\Delta;\emptyset]$ is the 2PI effective action of Cornwall, Jackiw and Tomboulis~\cite{Cornwall:1974vz}. The functionals $\Gamma[\phi;K]$ and $\Gamma[\Delta;J]$ are akin to the Routhian in classical mechanics in which not all sources are included in the Legendre transform. 

\begin{figure}[t!]
    \includegraphics[width=0.7\linewidth]{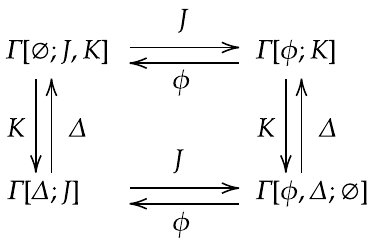}
    \caption{\label{fig:effectiveactions} The four potentials from \eqref{eq:fourpotentials} for the case $n=2$ and the Legendre transforms that connect them. The arrows indicate the coordinate with respect to which the Legendre transform is taken. Opposing sides of the square are connected by a double Legendre transform.}
\end{figure}

Figure~\ref{fig:effectiveactions} shows this family of four functionals. Each edge of this square represents a potential that depends on a different set of variables. The potentials on each edge are connected to nearby edges by single Legendre transforms, i.e., a Legendre transform with respect to one of $\phi$ or $\Delta$. This picture generalizes to larger $n$, although there are, in these cases, multiple Routhians that can be constructed by Legendre transforming with respect to only some of the sources in the Schwinger functional.

%%%%%%%%%%%%%%%%%%%%

\section{Associated Hessian manifolds}
\label{sec:hessman}

Consider a functional $\overline{\Gamma}$ that depends on $m$ sources $J^{(i)}$ and $n - m$ $i$-point functions $\Delta^{(i)}$. An example for this is the Schwinger functional depending on $n$ $i$-point sources and zero $i$-point functions. However, all following considerations can also be applied to functionals which have been obtained by Legendre transforms in the spirit of \cref{eqn:genEAA}. In any case, the configuration space is an infinite-dimensional manifold spanned by the $m$ sources and $n-m$ $i$-point functions. Each point is associated to a specific value of this set of $i$-point functions and sources. It will prove convenient to arrange these into a ``coordinate'' $\mathcal{Q}_A$, whose elements are of different dimensions. By convention, we choose this to be
\begin{equation}
    \mathcal{Q}_A=\begin{pmatrix} J^{(I_1)}\\ \vdots \\ J^{(I_m)}\\ \Delta^{(I^{\complement}_{1})}\\\vdots \\ \Delta^{(I^{\complement}_{n-m})}\end{pmatrix}\;,
    \label{eqn:Qcoordinates}
\end{equation}
where $I^{\complement}$ is the complement of $I$. The upper-case Latin indices are such that they run over all components of $\mathcal{Q}_A$. Without loss of generality, we assume that $\mathcal{Q}_A$ includes the one-point source $J^{(1)}$ in its first component. As we will see below, this assumption is still general enough to include all possible geometrical structures that can be constructed using a Hessian manifold. 

We start by equipping the configuration space with an affine connection $\bar{D}$. This connection is defined by the partial functional derivatives with respect to the  coordinates $\mathcal{Q}_A$; specifically,
\begin{equation}
    \bar{D}^A F_{B_1 \dots B_N}^{C_1 \dots C_N}[\mathcal{Q}_A] \equiv \fderiv{}{\mathcal{Q}_A} F_{B_1 \dots B_N}^{C_1 \dots C_N} [\mathcal{Q}_A]
    \label{eqn:Dbardef}
\end{equation}
for any tensor \smash{$F_{B_1 \dots B_N}^{C_1 \dots C_N}[\mathcal{Q}_A]$}. The coordinates $\mathcal{Q}_A$ form the affine coordinate frame of the connection $\bar{D}$. In any other coordinate frame, the $\bar{D}$ connection can obtain non-trivial Christoffel symbols, such that \cref{eqn:Dbardef} is modified.

With the introduction of the $\bar{D}$ connection, we have identified an affine structure on the configuration space. To identify the configuration space as a Hessian manifold, it is necessary to specify the metric. We define the inverse metric $g^{A B}$ as the Hessian of the functional $\overline{\Gamma}$ computed with the affine connection $\bar{D}$, as follows:
\begin{equation}
    g^{A B} = \bar{D}^A \bar{D}^B 
    \overline{\Gamma} \, .
\end{equation}
In the coordinate frame spanned by $\mathcal{Q}_A$, this can be expressed as
\begin{equation}\label{eq:gAB_up}
    g^{A B} = \fderiv{^2\overline{\Gamma}}{\mathcal{Q}_A \delta \mathcal{Q}_B} \, .
\end{equation}
With this choice for the metric, the configuration space becomes a Hessian manifold.

Note that the affine connection $\bar{D}$ is not metric compatible in general, viz.
\begin{equation}
    \bar{D}^A g^{B C} \neq 0 \, .
\end{equation}
However, there is a unique torsion-free and metric-compatible Levi--Civita connection $\nabla$. To find it, let us introduce the Amari--Chentsov tensor \cite{amari1980theory,amari1982differential,chentsov1982statiscal,nielsen2020elementary}
\begin{equation}
    \hat{\varGamma}^A_{B C} = \frac{1}{2} \bar{D}^A g_{B C} \, .
    \label{eqn:LCChristoffel2}
\end{equation}
It can be shown that \cref{eqn:LCChristoffel2} is fully symmetric once the index $A$ is lowered. This has wide-ranging implications, since any connection to which we add a term proportional to $\hat{\varGamma}^A_{B C}$ is a new connection. This allows us to define the so-called $\alpha$-connections \cite{nielsen2020elementary}
\begin{equation}
    \bar{D}^{(\alpha)} = \bar{D} + \alpha \hat{\varGamma} \otimes \bm{1} \, .
    \label{eqn:Dalphaconn}
\end{equation}
The tensor product denotes that $\hat{\varGamma}$ acts on all indices on which the derivative $\bar{D}^{(\alpha)}$ is acting~\cite{SHIMA1997277}. In essence, this shifts the Christoffel symbols included in $\bar{D}$ by $\alpha \hat{\varGamma}$.
The Levi--Civita connection is found from \cref{eqn:Dalphaconn} for $\alpha = -1$; that is,
\begin{equation}
    \nabla = \bar{D} - \hat{\varGamma}\otimes \bm{1} \, .
\end{equation}
This is the unique metric-compatible and torsion-free connection of the Hessian manifold. Note that, even though $\bar{D}$ is an affine connection, the Levi--Civita connection $\nabla$ is generally not flat and can give rise to a non-vanishing Riemann tensor.
Moreover, this is just one out of an infinite family of connections that is implied by Eq.~\cref{eqn:Dalphaconn}. 
Another important connection is given by
\begin{equation}
    D = \bar{D} - 2 \hat{\varGamma} \otimes \bm{1}\, .
    \label{eqn:Dconnection}
\end{equation}
In contrast to the Levi--Civita connection $\nabla$, it can be shown that $D$ is always affine, i.e., always a flat connection. Thus, we end up with a duality of affine connections:\ the $\bar{D}$ that we started with, and $D$ as defined in \cref{eqn:Dconnection}. In general, these are the only affine connections on a Hessian manifold. All other connections induced by \cref{eqn:Dalphaconn} are not affine and induce a non-vanishing Riemann tensor.

The affine coordinate frame of the $D$ connection is found from the gradient of the functional $\overline{\Gamma}$, i.e., 
\begin{equation}\label{eq:PA_def}
    \mathcal{P}^A \equiv \fderiv{
    \overline{\Gamma}}{\mathcal{Q}_A} \, .
\end{equation}
By noting that this relation is the familiar coordinate transformation induced by a Legendre transform, using \cref{eqn:Qcoordinates} in \cref{eq:PA_def}, we find that
\begin{equation}
    \mathcal{P}^A=\begin{pmatrix} \Delta^{(I_1)}\\ \vdots \\ \Delta^{(I_m)}\\ J^{(I^{\complement}_{1})}\\\vdots \\ J^{(I^{\complement}_{n-m})}\end{pmatrix}\;,
    \label{eqn:Pcoordinates}
\end{equation}
i.e., $i$-point sources $J^{(i)}$ are replaced by $i$-point functions $\Delta^{(i)}$ and vice versa. Thus, the affine coordinate frame of the  $\bar{D}$ connection  is given by the $\mathcal{Q}_A$ coordinates, and the affine coordinate frame of the $D$ connection is given by the $\mathcal{P}^A$ coordinates. In that sense, the coordinates $\mathcal{Q}_A$ and $\mathcal{P}^A$ both form affine coordinate frames, which are dual to each other.
This duality encoded by \cref{eq:PA_def} also leads to another relationship between the two affine structures $D$ and $\overline{D}$, which can be found using \cref{eq:gAB_up}:
\begin{align}
    \bar{D}^A&=\fderiv{}{\mathcal{Q}_A}=\fderiv{\mathcal{P}^B}{\mathcal{Q}_A}\fderiv{}{\mathcal{P}^B}=g^{AB}D_B\;.
\end{align}
Note that this relation is equivalent to \cref{eqn:Dconnection}.

We can express the metric as well as all other geometrical structures purely in terms of $\mathcal{P}^A$ instead of $\mathcal{Q}_A$. To do so, we have to introduce the dual potential, which we call $\Gamma$. It is related to the $\overline{\Gamma}$ functional by a Legendre transform with respect to all of its variables, such that
\begin{equation}
    \Gamma[\mathcal{P}
    ] = - \overline{\Gamma} [\mathcal{Q}]+ \mathcal{Q}_A \mathcal{P}^A \, .
    \label{eqn:gammadual}
\end{equation}
Note that the sum indicated by the repeated DeWitt indices runs over all components of the coordinates in \cref{eqn:Qcoordinates} and \cref{eqn:Pcoordinates}. It can then be shown that 
\begin{equation}
    g_{A B} = \fderiv{\Gamma}{\mathcal{P}^A \delta \mathcal{P}^B} \, ,
    \label{eqn:metricaffine}
\end{equation}
where we recall that the $\mathcal{P}^A$ are the affine coordinates of $D$, i.e.,
\begin{equation}
    D_A=\fderiv{}{\mathcal{P}^A}\;.
\end{equation}
Starting from this, we can also derive the Amari--Chentsov tensor \eqref{eqn:LCChristoffel2} and all other geometrical structures in terms of $\mathcal{P}^A$.

We are now in a position to justify our previous assumption of $\mathcal{Q}_A$ always containing the one-point source $J^{(1)}$ in \cref{eqn:Qcoordinates}. As we have seen, the geometry of the Hessian manifold always includes two affine coordinate sets, the $\mathcal{Q}_A$ coordinates that we started with in \cref{eqn:Qcoordinates}, and the $\mathcal{P}^A$ coordinates in \cref{eqn:Pcoordinates}. One of them always includes the one-point source $J^{(1)}$, while the other one contains the one-point function $\Delta^{(1)}$ instead. Which one of them we define to be $\mathcal{Q}_A$ does not alter the geometry. Interchanging $\mathcal{Q}_A$ with $\mathcal{P}^A$ leaves all distances unaffected. Using that such an interchange would also swap $\overline{\Gamma}$ with $\Gamma$, we can then show that
\begin{equation}
    \begin{split}
        \fderiv{\overline{\Gamma}}{\mathcal{Q}_A \delta \mathcal{Q}_B} \dd \mathcal{Q}_A \dd \mathcal{Q}_B =&\, \fderiv{\overline{\Gamma}}{\mathcal{Q}_A \delta \mathcal{Q}_B} \fderiv{\mathcal{Q}_A}{\mathcal{P}^C} \fderiv{\mathcal{Q}_B}{\mathcal{P}^D} \dd \mathcal{P}_C \dd \mathcal{P}_D \\
        =&\, \fderiv{\Gamma}{\mathcal{P}^A \delta \mathcal{P}^B} \dd \mathcal{P}^A \dd \mathcal{P}^B \, ,
    \end{split}
\end{equation}
where we have used that $\delta \mathcal{Q}_A / \delta \mathcal{P}^B = g_{A B}$. The left-hand side corresponds to a Hessian manifold constructed with initial coordinates including the one-point source, while the right-hand side corresponds to initial coordinates including the one-point function instead. Distances are the same in both cases.

Let us now consider $\overline{\Gamma} = \mathcal{W}$. In that case, we can compute $n$-point functions using the $\bar{D}$ derivative. This is because they are computed from the Schwinger functional by
\begin{equation}
    \langle \phi^{a_1} \dots \phi^{a_n} \rangle_{\text{c}} \equiv \fderiv{}{J_{a_1}} \dots \fderiv{}{J_{a_n}} \mathcal{W} \, .
    \label{eqn:npointdef2}
\end{equation}
Note that $J_a$ denotes the one-point source.
We can covariantize this by replacing functional derivatives w.r.t. the one-point source with the $\bar{D}$ derivative,
\begin{equation}
    \langle \phi^{a_1} \dots \phi^{a_n} \rangle_\text{c} = \bar{D}^{a_1} \dots \bar{D}^{a_n} \mathcal{W} \, .
    \label{eqn:npointcov}
\end{equation}
Here, we use the convention that geometrical objects evaluated with a lower-case Latin index give their one-point component, i.e.,
\begin{equation}
    \bar{D}^a = \fderiv{}{J_a} \, .
\end{equation}
Performing non-trivial transformations of $\mathcal{Q}_A$, the $\bar{D}$ derivatives pick up non-trivial Christoffel symbols that must be taken into account when working in other coordinates than $\mathcal{Q}_A$.

We remark that the connected $n$-point function from \eqref{eqn:npointcov} can be written in the form
\begin{equation}
        \langle \phi^{a_1} \dots \phi^{a_n} \rangle_\text{c} = \bar{D}^{a_1} \dots \bar{D}^{a_{n-2}} g^{a_{n-1}a_n} \, .
\end{equation}
Notice that the metric incompatibility of the connection $\bar{D}$ is, in fact, pivotal to us obtaining non-vanishing higher $n$-point functions. The corollary is that only theories with quadratic actions will have a metric-compatible $\bar{D}$. Identifying $g^{a_{n-1}a_n}=\Delta_{\rm c}^{a_{n-1}a_n}$ --- the connected two-point function --- and noting that $D$ naturally acts on $\Gamma[\mathcal{P};\emptyset]$, we can work entirely in terms of the coordinates $\phi^a$ and $\Delta_{\rm c}^{ab}=\Delta^{ab}-\phi^a\phi^b$. The connected $n$-point functions can then be expressed as
\begin{equation}
        \label{eq:connectedfunc}
        \langle \phi^{a_1} \dots \phi^{a_n} \rangle_\text{c} = \left[\prod_{i=1}^{n-2}g^{a_iB_i}D_{B_i}\right]\Delta^{a_{n-1}a_n}_{\rm c} \;.
\end{equation}
Note that, since we do not in general have metric compatibility of the connection $D$, the product in \eqref{eq:connectedfunc} is understood to be ordered as $g^{a_1B_1}D_{B_1}g^{a_2B_2}D_{B_2}\cdots g^{a_{n-2}B_{n-2}}D_{B_{n-2}}$. For the case $n = 2$, we write the operator $g^{aB}D_{B}$ explicitly in terms of the connected two-point function as
\begin{align}
    g^{aB}D_{B}&=\frac{\delta^2\mathcal{W}}{\delta J_a\delta J_b}\fderiv{}{\phi^b}+\frac{\delta^2\mathcal{W}}{\delta J_a\delta K_{\beta}}\fderiv{}{\Delta^{\beta}}\nonumber\\
    &=\Delta^{ab}_{\rm c}\left[\fderiv{}{\phi^b}-\frac{\delta^2 \Gamma}{\delta \phi^b\delta \Delta_{\rm c}^{\delta}}\left(\frac{\delta^2\Gamma}{\delta \Delta_{\rm c}^{\delta}\delta \Delta_{\rm c}^{\epsilon}}\right)^{-1}\fderiv{}{\Delta^{\epsilon}_{\rm c}}\right]\;.
\end{align}
This operator has been derived previously in the context of extracting $n$-point functions directly from the quantum effective action~\cite{VasilievKasanskii, Millington:2022cix}.

In summary, we have implemented the geometrical structure of a Hessian manifold in the configuration space of QFTs. Remarkably, this naturally implements a dual structure by which any potential $\overline{\Gamma}$ is related to a dual potential $\Gamma$ by a Legendre transform, see \cref{eqn:gammadual}. In the case that one of these potentials is the Schwinger functional, the dual potential is the full $n$PI effective action. However, we recall from \cref{fig:effectiveactions} that already at $n = 2$ there are four different potentials. These can be constructed by applying Legendre transforms with respect to the $i$-point sources/functions separately. Two of them are given by the Schwinger functional and the 2PI effective action, which form one Hessian manifold. The other two are given by the 1PI effective action and its dual potential, given by the double Legendre transform of the former. This leads to a second, different Hessian manifold. In general, since a Hessian manifold naturally includes a dual potential, each Hessian manifold involves a pair of potentials. For $n = 2$, this means that Hessian manifolds connect two sides of the square in \cref{fig:effectiveactions}.

%%%%%%%%%%%%%%%%%%%%

\section{2PI Effective Action as a Renormalization Group}
\label{sec:RG}

In this section, we focus on the Hessian manifold related to the Schwinger functional $\mathcal{W}[\emptyset;J,K]$ and the $2$PI effective action $\Gamma[\phi,\Delta;\emptyset]$. Moreover, we generalize to the case of $N$ fields, such that the DeWitt index in $\Phi^a$ becomes a multi-index and contains spacetime and internal indices of the field. We have
\begin{equation}
    \mathcal{Q}_A=\begin{pmatrix} J_{a}\\ K_\alpha\end{pmatrix}\;,\quad\text{and}\quad \mathcal{P}^A=\begin{pmatrix} \phi^a \\ \Delta^\alpha \end{pmatrix}
    \label{eqn:2PIaffcoordinates}
\end{equation}
As before, the upper-case Latin indices are such that they run first over $J_a$ or $\phi^a$, and then over $K_\alpha$ or $\Delta^\alpha$. Moreover, if we evaluate a coordinate with a lower-case Latin index, we get back a one-point object, i.e.,
\begin{equation}
    \mathcal{Q}_a = J_a \, , \qquad \mathcal{P}^a = \phi^a \, .
\end{equation}
Similarly, we employ the convention introduced earlier that lower-case Greek letters correspond to two-point objects, i.e.,
\begin{equation}
    \mathcal{Q}_\alpha = K_\alpha = K_{a b} \, , \qquad \mathcal{P}^\alpha = \Delta^\alpha = \Delta^{a b} \, .
\end{equation}

We will now interpret the source for the two-point function $K_\alpha$ as a regulator term. 
This means, we specify the form of $K_\alpha$ such that it can be interpreted as a regulator, either to regularize UV or IR modes. The Schwinger functional and the quantum effective action then form a RG, the renormalization scale $k$ is in direct relation to the regulator $K_\alpha$.

If the regulator is chosen to regularize UV modes, it serves as a cutoff such that modes in the deep UV are cut out from the path integral. To do so, the regulator $K_\alpha$ must diverge in the UV, with
\begin{equation}
    \frac{1}{K_\alpha} \to 0 \quad \text{for} \quad \frac{p^2}{k^2} \to \infty \, ,
\end{equation}
where $p^2$ is the momentum of the fluctuation modes in the path integral.
In the IR, a UV cutoff should not induce any effects. Hence, it must fall off for small momenta, with
\begin{equation}
    K_\alpha \to 0 \quad \text{for} \quad \frac{p^2}{k^2} \to 0 \, .
\end{equation}
If both criteria are fulfilled, we can interpret $K_\alpha$ as a UV cutoff for the Schwinger functional. Changing the renormalization scale $k$ then induces a flow for the Schwinger functional. This can be encoded in a FRG for the effective action, such as the Polchinski equation~\cite{Polchinski:1983gv}. 

For many practical purposes, it is useful to consider a flow equation for the quantum effective action instead of the Schwinger functional. This can be achieved by introducing an IR regulator into the Schwinger functional. Such a regulator should induce a mass $k^2$ for low IR modes. Thus, for an IR regulator we require
\begin{equation}
    K_\alpha \to k^2 \quad \text{for} \quad p^2/k^2 \to 0 \, .
\end{equation}
UV modes should be unaffected by an IR regulator, leading to the second condition that
\begin{equation}
    K_\alpha \to 0 \quad \text{for} \quad p^2/k^2 \to \infty \, .
\end{equation}
The RG flow encoding the change of the effective action when changing the RG scale is given by the Wetterich equation~\cite{Wetterich:1992yh, Morris:1993qb, Ellwanger:1993mw}. In practical calculations, the Wetterich flow often has favourable properties in terms of convergence when compared to the Polchinski flow.

The RG equations for both, the Polshinski and the Wetterich flow, are well-understood and can be applied to perform practical computations. Here, we aim to understand the RG flow in terms of geometrical structures provided by the Hessian manifold. For the most part, we focus on the flow for the quantum effective action. 

As discussed above, starting from the Schwinger functional the Hessian manifold implies a natural dual functional, which is given by the 2PI effective action in terms of the variables $\phi^a$ and $\Delta^\alpha$. To view this in terms of a RG, we interpret $\Delta^\alpha$ as a functional of $\phi^a$ and $K_\alpha$,
\begin{equation}
    \Delta^\alpha = \Delta^\alpha \left[ \phi, K \right] \, .
    \label{eqn:RGDelta}
\end{equation}
The flow of the so-obtained $k$-dependent effective action can be given in the implicit form~\cite{Alexander:2019cgw}
\begin{equation}
    \partial_t \Gamma = \fderiv{\Gamma}{\Delta^\alpha} \partial_t \Delta^\alpha = K_\alpha \partial_t \Delta^\alpha \, ,
    \label{eqn:2PIflow}
\end{equation}
where the $t$-derivative is understood as varying $K_\alpha$ while keeping $\phi^a$ fixed. Due to this, \cref{eqn:2PIflow} is only implicit since additional information about the relation of $\Delta^\alpha$ to $K_\alpha$ is required in order to close the equation. This can be provided using the convexity of the effective action, as described in App.~\ref{sec:closure}. 

The RG is then a coordinate transformation for the 2PI effective action. Keeping $\phi^a$ constant, different values for the RG scale $k$ imply different values for the two-point function $\Delta^\alpha$ via \cref{eqn:RGDelta}. Thus, changing $k$ and keeping $\phi^a$ fixed leads to a vector field in configuration space which encodes the RG evolution. Complementary to this vector field there are hypersurfaces of constant RG scale $k$.
Every point on such a hypersurface is associated to the same value for the source $K_\alpha$ but varies in $\phi^a$. The vector field generated by the RG evolution connects hypersurfaces with different RG scales. Below, we will see that both of these geometrical structures have a natural identification in terms of the geometry of a Hessian manifold. 

%%%%%%%%%%%%%%%%%%%%

\subsection{Renormalization Group as Geodesics}

The basic idea to understand the RG in terms of geometrical structures is to construct geodesics whose proper time $t$ is in a linear relation with the two-point source $K_\alpha$. Then, starting from any point in configuration space with given $K_\alpha$, such a geodesic moves to a different point in configuration space, changing the value of $K_\alpha$ linearly. 

If we require the change of $K_\alpha$ to be constant along a trajectory, we must have
\begin{equation}
    \dderiv{K_\alpha}{t} = \text{const} \, ,
    \label{eqn:Kconst}
\end{equation}
where $t$ is the proper time parameter of the trajectory. The implications of this equation can be seen by using $K_\alpha = \fderiv{\Gamma}{\Delta^\alpha}$ and taking a second derivative of \cref{eqn:Kconst}. Taking note of (\ref{eqn:metricaffine}) and (\ref{eqn:LCChristoffel2}), this yields the requirement
\begin{equation}
    g_{\alpha B} \left[ \dderiv{\mathcal{P}^B}{t^2} + 2 \varGamma^{B}_{C D} \dderiv{\mathcal{P}^C}{t} \dderiv{\mathcal{P}^D}{t} \right] = 0 \, .
    \label{eqn:constKreq}
\end{equation}
Recalling \cref{eqn:Dconnection}, we see that the factor of 2 in front of the Levi--Civita Christoffel symbols $\varGamma^{B}_{C D}$ tells us that the equation in the brackets is the geodesic equation for the connection $\bar{D}$.\footnote{This result can also be obtained more directly by noting that \cref{eqn:constKreq} implies a straight line in the affine coordinates of $\bar{D}$. Thus, it must be a geodesic of the $\bar{D}$ connection.}  Thus, requiring the trajectory to have constant change in $K_\alpha$, such a trajectory must either be a geodesic of the connection $\bar{D}$, or the geodesic equation must be fulfilled when multiplying with the metric.

In the following, we assume that the trajectory is indeed a solution for the $\bar{D}$ geodesics. Thus,
\begin{equation}
    \dderiv{\mathcal{P}^A}{t^2} + 2 \varGamma^{A}_{B C} \dderiv{\mathcal{P}^B}{t} \dderiv{\mathcal{P}^C}{t} = 0 \, .
    \label{eqn:Dbargeo}
\end{equation}
Any trajectory subject to \cref{eqn:Dbargeo} leads to a constant change of $K_\alpha$. The same holds equivalently for $J_a$, as both $\dot{K}_\alpha$ and $\dot{J}_a$ are conserved quantities along a $\bar{D}$ geodesic. The initial values for $K_\alpha$ and $J_a$ as well as their derivatives can be chosen as integration constants at an initial scale of the geodesic.

The properties explained above are very useful when thinking in terms of an RG evolution. In the following, we will see how the $\bar{D}$ geodesics can be used
\begin{itemize}
    \item to construct surfaces of constant RG time.
    \item to generate an RG flow.
\end{itemize}

%%%%%%%%%%%%%%%%%%%%%%%%%%%%%%

\subsection{Surfaces of Constant RG Scale}
\label{sec:constrg}

Any trajectory within a hypersurface of constant RG scale $k$ must have the same value for the RG scale everywhere by definition. Thus, if we parametrize such surfaces in terms of trajectories, we require that
\begin{equation}
    \dderiv{K_\alpha}{t} = 0
    \label{eqn:Kfixed}
\end{equation}
everywhere on such a trajectory. These trajectories are found by taking the $\bar{D}$ geodesics and imposing \cref{eqn:Kfixed} as an initial condition. Due to the properties of the $\bar{D}$ connection, \cref{eqn:Kfixed} is conserved along the whole geodesic. Surfaces of constant $K_\alpha$ can then be constructed by viewing all $\bar{D}$ geodesics with initial condition \cref{eqn:Kfixed} together. 

While the initial condition \cref{eqn:Kfixed} is a requirement to lead to constant $K_\alpha$, the remaining initial conditions can be used to parametrize the hypersurface. The value of $K_\alpha$ specifies the RG scale $k$. The values for $J_a$ and $\dot{J}_a$ specify the value of $\phi^a$ on the trajectory and how it changes with the proper time parameter $t$. Note that the relation between $J_a$ and $\phi^a$ is $K_\alpha$ dependent and, in general, non-linear. 

\begin{figure}[t!]
    \includegraphics[width=0.99\linewidth]{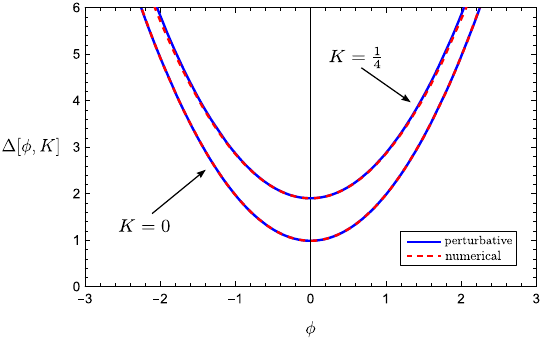}
    \caption{Surfaces of constant $K$ are shown for the zero-dimensional $\phi^4$ theory. Bottom lines show the surface at $K = 0$ and upper lines at $K = 1/4$. Blue lines show the exact result at second order in the coupling. Red dashed lines correspond to the solution of numerically integrating the geodesic equation.}
    \label{fig:constK}
\end{figure}

Let us discuss in some more detail how trajectories subject to \cref{eqn:Kfixed} can be constructed within the coordinate frame spanned by $\phi^a$ and $\Delta^\alpha$. The first task is to translate \cref{eqn:Kfixed} into the $\phi^a$ and $\Delta^\alpha$ frame such that it can be used as an initial condition for the geodesic. Using $K_\alpha = \fderiv{\Gamma}{\Delta^\alpha}$, we find
\begin{equation}
    g_{\alpha A} \dderiv{\mathcal{P}^A}{t} = g_{\alpha a} \dderiv{\phi^a}{t} + g_{\alpha \beta} \dderiv{\Delta^\beta}{t} = 0 \, .
    \label{eqn:constKinitial}
\end{equation}
While \cref{eqn:constKinitial} gives the initial condition to fulfil $\dot{K}_\alpha=0$, the value for $K_\alpha$ associated with that trajectory has to be obtained by evaluating
\begin{equation}
    \fderiv{\Gamma}{\Delta^\alpha} = K_\alpha \, .
    \label{eqn:K}
\end{equation}
Since this is constant by construction, it is sufficient to evaluate \cref{eqn:K} at the initial point of the trajectory. Together with the remaining initial conditions, the solutions of \cref{eqn:Dbargeo} then allow us to parametrize complete surfaces of constant $K_\alpha$. 

As an example for the considerations above, we consider the zero-dimensional QFT of a scalar field $\phi$ with a $\phi^4$ interaction. Zero-dimensional theories have proved useful elsewhere for illustrating the properties of the effective action and the FRG~\cite{Millington:2019nkw, Millington:2021ftp, Koenigstein:2021syz, Koenigstein:2021rxj, Steil:2021cbu, Millington:2022cix}. The classical action is given by
\begin{equation}
    S =  \frac{1}{2} \phi^2 + \frac{\lambda}{4!} \phi^4 \,.
    \label{eqn:0dimqft}
\end{equation}
For this simple case, the Schwinger functional can be computed explicitly as a perturbative expansion in $\lambda$, giving
\begin{equation}
    \begin{split}
        \mathcal{W}[J, K] =&\, \ln \int \dd \phi\; e^{-S + J \phi + K \phi^2} \\
        =&\, \frac{1}{2} \left(\frac{J^2}{\tilde{K}}-\ln \tilde{K} +\ln (2 \pi )\right) \\
        & - \lambda \frac{J^4+6J^2 \tilde{K}+3 \tilde{K}^2}{24 \tilde{K}^4}
         + \order{\lambda^2} \, ,
    \end{split}
    \label{eqn:Wsol}
\end{equation}
where we have defined $\tilde{K} = 1 - 2K$.
With this to hand, all geometrical quantities and the effective action can be given as an explicit perturbative expansion in $\lambda$ as well.

To construct geodesics of the $\bar{D}$ connection, we can start by deriving expressions for $\phi$ and $\Delta$ in terms of $J$ and $K$. These are obtained by taking derivatives of $W$ with respect to $J$ and $K$, yielding
\begin{equation}
    \begin{split}
        \phi =&\, \frac{J}{\tilde{K}} - \lambda \frac{J^3+3 J \tilde{K}}{6 \tilde{K}^4}
         + \order{\lambda^2} \, , \\
        \Delta =&\, \frac{J^2+\tilde{K}}{\tilde{K}^2}-\frac{\lambda  \left(2 J^4+9 J^2 \tilde{K}+3 \tilde{K}^2\right)}{6 \tilde{K}^5} \\
        & + \order{\lambda^2} \, .
    \end{split}
    \label{eqn:phidelta}
\end{equation}
While this is the general relation for $\phi$ and $\Delta$ in terms of $J$ and $K$, we can easily transform it into an expression for a $\bar{D}$ geodesic. For this, we note that $\bar{D}$ geodesics are straight lines in $J$ and $K$. Thus, we must have
\begin{equation}
    \begin{split}
        J =&\, J_0 + \dot{J}_0 t \, , \\
        K =&\, K_0 + \dot{K}_0 t \, ,
    \end{split}
    \label{eqn:Dbargeosol}
\end{equation}
with $J_0$, $\dot{J}_0$, $K_0$, and $\dot{K}_0$ the open parameters (initial conditions) of the geodesic. Note that these are not straight lines anymore in the $\phi$ and $\Delta$ coordinates. This is due to the fact that the relationship between the sources and expectation values is non-linear, see \cref{eqn:phidelta}. Thus, the $\bar{D}$ geodesics are straight lines in $J$ and $K$, but not in $\phi$ and $\Delta$.

While, for the simple case of a zero-dimensional QFT field, we can explicitly obtain the relations between $\phi$ and $\Delta$, the same equations can also be obtained by working directly in the $\phi$ and $\Delta$ coordinates following the geodesic equation \cref{eqn:Dbargeo} with appropriate initial conditions. In our example, these geodesic equations reduce to two coupled differential equations for $\phi$ and $\Delta$ of the form
\begin{equation}
    \begin{split}
        0 =&\, \ddot{\phi} + 2 \dot{\phi} \frac{2 \phi \dot{\phi} - \dot{\Delta}}{\Delta - \phi^2} \\
        & + \lambda  \left[-4 \phi^2 \dot{\Delta} \dot{\phi} + \phi \dot{\Delta}^2+\left(5 \phi^3 - \Delta \phi \right) \dot{\phi}^2\right] \\
        & + \order{\lambda^2} \, , \\
        0=&\, \ddot{\Delta} + 2 \dot{\Delta} \frac{2 \phi \dot{\phi} - \dot{\Delta}}{\Delta - \phi^2} - 2 \dot{\phi}^2 + \\
        & \lambda  \big[-2 \phi \left(\Delta + 3 \phi^2\right) \dot{\Delta} \dot{\phi} + \left(\Delta + \phi^2\right) \dot{\Delta}^2 \\
        & +\left(-4 \Delta \phi^2+\Delta^2 + 11 \phi^4\right) \dot{\phi}^2\big] + \order{\lambda^2} \, .
    \end{split}
\end{equation}
It can be checked explicitly that the general solution is given by \cref{eqn:phidelta}, with \cref{eqn:Dbargeosol}. In more complicated theories where analytical solutions are not available, such equations can equally well be solved using numerical integration once the initial conditions are specified.

To set the initial conditions, we note that the parameter $K_0$ is given by the RG scale of the surface, and we must choose $\dot{K}_0 = 0$ to ensure that the trajectory indeed gives rise to a constant RG scale.
The remaining initial conditions $J_0$ and $\dot{J}_0$ can be chosen arbitrarily as long as $\dot{J}_0 \neq 0$. While, in higher dimensions, these parametrize different directions on a surface of constant RG scale, in the case of a zero-dimensional QFT with one scalar field, such surfaces are one dimensional. Thus, different choices for $J_0$ and $\dot{J}_0$ lead to the same trajectory.

In \cref{fig:constK}, we show RG surfaces at $K = 0$ and $K = 1/4$ with $\lambda = 3/100$. The blue lines show the known results obtained at second order in $\lambda$, i.e., \cref{eqn:phidelta} with \cref{eqn:Dbargeosol}. The dashed red line corresponds to the result of solving the geodesic equation \cref{eqn:Dbargeo} with initial conditions set to fulfil \cref{eqn:constKinitial} and \cref{eqn:K}. Both lines fully agree up to higher-order effects in the coupling, which become visible for large values of $\phi$.

%%%%%%%%%%%%%%%%%%%%

\subsection{RG Flow}
\label{sec:rgflow}

In this section, we show how the RG evolution can be implemented using the $\bar{D}$ geodesics. The idea is very similar to the construction of the previous section, however, here we choose the derivative of $K_\alpha$ with respect to the proper time to be non-vanishing and constant, i.e., 
\begin{equation}
    \dderiv{K_\alpha}{t} = F_\alpha = \text{const} \, .
    \label{eqn:Knonvan}
\end{equation}
With \cref{eqn:Knonvan} as an initial condition, $\bar{D}$ geodesics generate an RG flow. Starting from any point where the regulator takes the value $K_\alpha$, this geodesic connects to a point with the regulator equal to $K_\alpha + t F_\alpha$. Note that the value for the one-point function $\phi^a$ is not conserved along this trajectory and changes. This is because the $\bar{D}$ geodesics conserve $\dot{J}_a$ and $\dot{K}_\alpha$ but not necessarily $\dot{\phi}^a$ or $\phi^a$. Since the relationship between $J_a$ and $\phi^a$ is non-linear in general, it is difficult to deduce general statements on the variation of $\phi^a$ along a $\bar{D}$ geodesic. However, the crucial point is that the change of $K_\alpha$ along the geodesic is independent of the position in configuration space. It only depends on $K_\alpha$ at the initial point and the value for the proper time. Hence, starting from any point on the surface with regulator $K_\alpha$, we must end up on a surface with regulator $K_\alpha + t F_\alpha$. Even though the value of $\phi^a$ changes along the trajectory, we can reconstruct the full surface at $K_\alpha + t F_\alpha$ by considering the whole surface at $K_\alpha$ and evolving each point along the $\bar{D}$ geodesic.

In practice, we can implement \cref{eqn:Knonvan} as a condition in the $\phi^a$, $\Delta^\alpha$ coordinates as well. The initial condition \cref{eqn:Knonvan} translates to
\begin{equation}
    g_{\alpha b} \dderiv{\phi^b}{t} + g_{\alpha \beta} \dderiv{\Delta^\beta}{t} = F_\alpha \, .
    \label{eqn:geocon}
\end{equation}
One of $\dot{\phi}^b$ and $\dot{\Delta}^\beta$ can be chosen freely. We may choose $\dot{\phi}^b = 0$, however, note that this is only fulfilled at the initial point of the $\bar{D}$ geodesic and not for later proper times. This is due to the fact that $\dot{\phi}^b = 0$ is not a conserved quantity along the $\bar{D}$ geodesics.

\begin{figure}[t!]
    \includegraphics[width=0.99\linewidth]{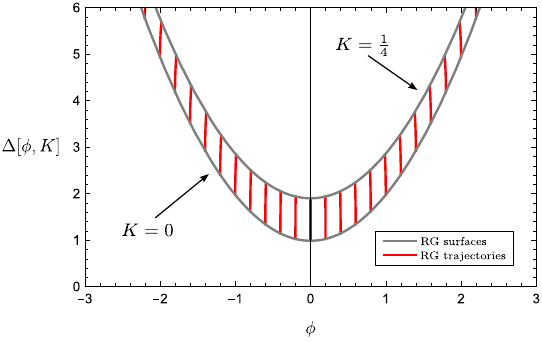}
    \caption{The RG evolution as generated by $\bar{D}$ geodesics is shown for a zero-dimensional $\phi^4$ theory. The grey lines indicate the surfaces of constant RG time at $K = 0$ and $K = 1/4$. Red lines show $\bar{D}$ geodesics evolving each point from $K = 0$ to one point on the $K = 1/4$ surface.}
    \label{fig:RGevol}
\end{figure}

Let us discuss the implementation of this idea for a single scalar field  in a zero-dimensional $\phi^4$ theory, i.e., \cref{eqn:0dimqft}. Note that our considerations are readily extended to more complicated fields, e.g., $N$ scalar fields, using appropriate indices in the following expressions. We take $F = 1$ and $\dot{\phi} = 0$ as initial conditions. The initial condition for $\dot{\Delta}$ is inferred from \cref{eqn:geocon}. We then start from a given surface at $K = 0$ on which $\Delta$ is fixed as a function of $\phi$. For each point on this surface, we construct a geodesic and evolve it to proper time $t = 1/4$. This can be done either numerically or analytically as a perturbative expansion in the coupling.\footnote{To obtain analytical results, it is most convenient to work in the $\{J$, $K\}$ coordinates where the $\bar{D}$ geodesics correspond to straight lines. Using perturbative relations between the sources and expectation values, we then obtain the geodesics in the $\{\phi$, $\Delta\}$ frame.} The result is shown in \cref{fig:RGevol}. At the end of each trajectory, we end up with $K = 1/4$ forming the surface of $K = 1/4$. Due to our choice of initial conditions, the value for $\phi$ is almost constant along the geodesics. However, small variations are present when zoomed into the graph. These changes in $\phi$ are not originating from neglecting higher orders in the perturbative expansion. This can be seen from the analytical solution of $\phi$ along this family of geodesics
\begin{equation}
        \phi = \phi_\text{ini} - \lambda \frac{2 t^2 \phi_\text{ini}}{(1-2 t)^2}+O\left(\lambda ^2\right) \, .
    \label{eqn:phiconstsol}
\end{equation}
Already at first order in the coupling, there is a non-trivial $t$-dependence. Thanks to our chosen initial condition, this starts at second order in $t$ and the effect is rather small. Note that the effect is absent altogether for $\phi_\text{ini} = 0$.

Let us now switch back to the general case of a finite-dimensional QFT. We might be wondering whether there can be a trajectory with a parameter $t$ such that the derivative of the two-point source $K_\alpha$ by $t$ is constant along the trajectory and the derivative of the one-point function $\phi^a$ vanishes for all $t$, i.e.,
\begin{equation}
    \dderiv{\phi^a}{t} = 0 \, .
    \label{eqn:phiconst}
\end{equation}
We can translate the latter into the source coordinates. Using $\phi^a = \fderiv{\mathcal{W}}{J_a}$ and $g^{A B} = \fderiv{^2 \mathcal{W}}{\mathcal{J}_A \delta \mathcal{J}_B}$, we have
\begin{equation}
    g^{a b} \dderiv{J_b}{t} + g^{a \beta} \dderiv{K_\beta}{t} = 0 \, .
\end{equation}
Now, we use the second requirement of keeping $\dot{K}_\beta = F_\beta = \text{const}$ along the trajectory. Thus, we must fulfil the relation
\begin{equation}
    g^{a b} \dderiv{J_b}{t} = - g^{a \beta} F_{\beta} \, .
    \label{eqn:phiconstKt}
\end{equation}
In general, there is no reason for this to have a solution with $\dot{J}_b = \text{const}$ in the interacting case.\footnote{We have explicitly checked that \cref{eqn:phiconstKt} is violated in a zero-dimensional $\phi^4$ theory.} Thus, constructing a trajectory with $\dot{K}_\beta = \text{const}$ and $\dot{\phi}^a = 0$, generally leads to $\ddot{J}_a \neq 0$, which means that such a trajectory cannot be a geodesic of $\bar{D}$, and \cref{eqn:Dbargeo} is violated. However, the more general condition \cref{eqn:constKreq} must still be fulfilled since it is a direct consequence for any trajectory with $\dot{K}_\alpha = \text{const}$. 
In fact, trajectories with the properties \cref{eqn:phiconst} and \cref{eqn:Knonvan} do exist. One way to find such trajectories is by using the 2PI FRG~\cite{Alexander:2019cgw, Millington:2021ftp}. However, it is unclear how the geometrical meaning of those trajectories is related to the Hessian manifold analysed here.

%%%%%%%%%%%%%%%%%%%%

\subsection{1PI Effective Action}

The discussion above shows that we can find geodesics on the Hessian manifold introduced with the affine coordinates \cref{eqn:2PIaffcoordinates} to implement an RG flow. However, these geodesics do not keep the one-point function $\phi^a$ invariant. Starting from a given point in the configuration space, the RG flow will generally end up at a different value for $\phi^a$. Although we can formally define trajectories with a constant change of  $K_\alpha$ while keeping $\phi^a$ constant, their interpretation in terms of geometric quantities of the Hessian manifold as set up with \cref{eqn:2PIaffcoordinates} is unclear. As we will show here, a geometrical interpretation of such trajectories can be provided by constructing a different Hessian manifold starting from the 1PI effective action $\Gamma[\phi;K]$.\footnote{Note that this is the usual 1PI effective action in the presence of a two-point source $K_\alpha$.} Its natural variables are $\phi^a$ and $K_\alpha$. This is the key difference to the 2PI effective action introduced above. The dependence on the two-point function $\Delta^\alpha$ has been traded for a dependence on the two-point source $K_\alpha$ by a Legendre transform; specifically,
\begin{equation}
    \begin{split}
        \Gamma[\phi;K] =&\, \Gamma[\phi,\Delta;\emptyset] - \Delta^\alpha K_\alpha \\
        =&\, -\mathcal{W}[J,K] + J_a \phi^a \, .
    \end{split}
    \label{eqn:1PI2PI}
\end{equation}
Instead of defining a Hessian manifold in the sense of \cref{eqn:2PIaffcoordinates}, we can now equally well define a Hessian manifold with \cref{eqn:1PI2PI} as a starting point. In this case, the $D$ connection, which we call $D^\text{1PI}$ to avoid confusion with the connection introduced above, is defined by the partial derivatives in the $\{\phi^a$, $K_\alpha\}$ frame, and the metric is defined by the Hessian of $\Gamma[\phi;K]$ using the $D^\text{1PI}$ connection.

The construction \cref{eqn:1PI2PI} has the advantage that geodesics of the resulting $D^\text{1PI}$ connection are straight lines in $\{\phi^a$, $K_\alpha\}$. Thus,
\begin{equation}
    \dot{\phi}^a = \text{const} \, , \quad \dot{K}_\alpha = \text{const} \, .
\end{equation}
Implementing the RG using the same concepts as in \cref{sec:constrg} and \cref{sec:rgflow}, the RG flow induced by the $D^\text{1PI}$ geodesics can be chosen to keep $\phi^a$ constant, while $K_\alpha$ experiences a constant change. The price to pay for this construction is a more involved relation to the Schwinger functional. While the Schwinger functional of the Hessian manifold in \cref{sec:hessman} is a potential for the metric, this is not the case if we construct a Hessian manifold starting from \cref{eqn:1PI2PI}. The dual potential of \cref{eqn:1PI2PI} is given by the Legendre transform with respect to all of its variables. The variables dual to $\phi^a$ and $K^\alpha$ are found by the derivatives of $\Gamma[\phi;K]$
\begin{equation}
    J_a = \fderiv{\Gamma[\phi;K]}{\phi^a} \, , \qquad \tilde{\Delta}^\alpha \equiv - \Delta^\alpha = \fderiv{\Gamma[\phi;K]}{K_\alpha} \, .
    \label{eqn:1PIvars}
\end{equation}
Employing the Legendre transform, we find the dual potential to be given by
\begin{equation}
    \begin{split}
        \Gamma[\Delta;J] &=\, - \Gamma[\phi;K] + K_\alpha \tilde{\Delta}^\alpha + J_a \phi^a \\
        &= \mathcal{W}[J,K] + K_\alpha \tilde{\Delta}^\alpha \, .
    \end{split}
    \label{eqn:1PIdual}
\end{equation}
Due to the minus sign in the definition of $\tilde{\Delta}^\alpha$ in \cref{eqn:1PIvars}, this is a Legendre transform of the Schwinger functional. Thus, in contrast to the construction above, the Schwinger functional does not arise as one of the potentials for the Hessian manifold anymore, but only its Legendre transform.

Despite the differences between the Hessian manifolds constructed from \cref{eqn:2PIaffcoordinates} and using \cref{eqn:1PI2PI}, the RG flows are equivalent. In both cases we can construct surfaces with $K_\alpha = \text{const}$. Since such surfaces are uniquely given by a value for $K_\alpha$, they must be the same in both cases. The RG flow between them, as induced by the geodesics of $\bar{D}$ or $D^\text{1PI}$ only, differs in what variables are kept constant. The former can generally not be chosen to keep $\dot{\phi}^a = \text{const}$ (see \cref{eqn:phiconstsol}), while the latter does. However, the RGs obtained from both are equivalent and contain the same information.

%%%%%%%%%%%%%%%%%%%%

\section{Conclusions}
\label{sec:conc}

In this work, we have implemented the non-perturbative effective actions of quantum field theory by means of the geometrical structures of Hessian manifolds. By associating the metric of the configuration space with the Hessian of the effective action, a natural relationship between the Schwinger functional and the effective action is established through dual potentials of the Hessian structure. This duality extends to sources and $i$-point functions, which arise as two sets of affine coordinates related by a Legendre transform. The partial derivatives of these affine coordinates establish two affine connections on the manifold.

By extending these insights to the FRG, working within the 2PI formulation unveils novel interpretations of the RG in terms of these geometric structures. The origin of this lies in the fact that RG transformations in this framework correspond to coordinate changes in configuration space. Surfaces of constant RG scale manifest as fixed-value surfaces for the two-point source. Since the value for the two-point source is related to an affine structure, geodesics can be constructed whose proper-time parameters have specific properties with respect to the RG scale. First, geodesics can be constructed that keep the RG scale constant. These can be used to reconstruct whole surfaces of constant RG scale. Second, geodesics can be found whose proper-time parameters are in a one-to-one correspondence with the RG scale. These can be used to implement RG transformations via geodesics. Thus, we find that RG information is naturally encoded in geodesics of the Hessian manifold of the $2$PI effective action.

We leave for further work the generalisation of these geodesic flows beyond the case of $n=2$ and to higher dimensional configuration spaces, involving higher $i$-point correlation functions.

%%%%%%%%%%%%%%%%%%%%

\begin{acknowledgments}
This work was supported by a United Kingdom Research and Innovation (UKRI) Future Leaders Fellowship [Grant No.~MR/V021974/2] and by the Science and Technology Facilities Council (STFC) [Grant Nos.~ST/T000732/1 and~ST/X00077X/1]. YK is grateful for the hospitality of Perimeter Institute where part of this work was carried out. Research at Perimeter Institute is supported in part by the Government of Canada through the Department of Innovation, Science and Economic Development and by the Province of Ontario through the Ministry of Colleges and Universities. For the purpose of open access, the authors have applied a Creative Commons Attribution (CC BY) licence to any Author Accepted Manuscript version arising.
\end{acknowledgments}

%%%%%%%%%%%%%%%%%%%%

\section*{Data Access Statement}

No data were created or analysed in this study.

%%%%%%%%%%%%%%%%%%%%

\appendix

%%%%%%%%%%%%%%%%%%%%

\section{Hesse--Koszul Flow}
\label{sec:HesseKoszul}

In~\cite{MIRGHAFOURI201754, ARXIV.2001.02940}, a geometrical flow for the metric was constructed, which preserves its Hessian property. This was achieved using the second Koszul form
\begin{equation}
    \beta_{a b} = \frac{1}{2} \fderiv{^2}{\phi^a \delta \phi^b} \ln \det g_{c d} = \frac{1}{2} \fderiv{^2}{\phi^a \delta \phi^b} \text{Tr} \ln{g_{c d}} \, ,
\end{equation}
and the flow was defined by\footnote{Compared to \cite{MIRGHAFOURI201754}, this expression differs by a factor of $2$, which has been absorbed into the definition of the flow time $t$.}
\begin{equation}
    \partial_t g_{a b} = \beta_{a b} \, .
    \label{eqn:KoszulFlow}
\end{equation}
Commuting the functional derivatives with the functional trace, we can write this flow in the form
\begin{equation}
        \beta_{a b} = - \frac{1}{2} g^{c d} \fderiv{g_{d e}}{\phi^a} g^{e f} \fderiv{g_{f c}}{\phi^b} + \frac{1}{2} g^{c d} \fderiv{^2g_{c d}}{\phi^a \delta \phi^b}\,.
\end{equation}
This has some structural similarity with the flow of the two-point function derived from the Wetterich equation
\begin{equation}
    \partial_t \Gamma_k = \frac{1}{2} \left( \partial_t \mathcal{R}_k \right)_{a b} \mathcal{G}^{b a}\,,
\end{equation}
with
\begin{equation}
    \mathcal{G}_{a b} = g_{a b} + \mathcal{R}_{k, a b} \, ,
\end{equation}
and $\mathcal{G}^{a b}$ its inverse. $\Gamma_k$ is the so-called 1PI effective average action~\cite{Tetradis:1993ts}. Taking two functional derivatives, noting that $\mathcal{R}_k$ is independent of $\phi^a$, this gives
\begin{equation}
    \begin{split}
        \partial_t g_{a b} =& \, 2 \left( \partial_t \mathcal{R}_k \right)_{c d} \mathcal{G}^{d e} \fderiv{\mathcal{G}_{e f}}{\phi^a} \mathcal{G}^{f g} \fderiv{\mathcal{G}_{g h}}{\phi^b} \mathcal{G}^{h c} \\
        &- \left( \partial_t \mathcal{R}_k \right)_{c d} \mathcal{G}^{d e} \fderiv{^2 \mathcal{G}_{e f}}{\phi^a \delta \phi^b} \mathcal{G}^{f c} \, .
    \end{split}
\end{equation}
Note also that the expression for the one-loop effective action is equivalent to the second Koszul form if we take two functional derivatives.

Instead of Eq.~\cref{eqn:KoszulFlow}, we might also consider a flow for the potential itself using the Monge--Amp\'ere operator~\cite{ARXIV.2001.02940}
\begin{equation}
    \mathcal{M} [\Gamma] \equiv \det g_{a b} \, .
\end{equation}
The Hesse--Koszul flow is then represented as
\begin{equation}
    \partial_t \Gamma = \frac{1}{2} \ln \mathcal{M} [\Gamma] = \frac{1}{2} \text{Tr} \ln g_{a b} = \frac{1}{2} \text{Tr} \ln \fderiv{\Gamma}{\phi^a \delta \phi^b} \, ,
\end{equation}
wherein we note the structural similarity to the one-loop effective action
\begin{equation}
    \Gamma^{(1)} = \frac{1}{2} \text{Tr} \ln \fderiv{S}{\phi^a \delta \phi^b} \, .
\end{equation}

%%%%%%%%%%%%%%%%%%%%

\section{Closing the 2PI RG}
\label{sec:closure}

In this appendix, we review the closure of the 2PI flow equation based on convexity. We take $\Gamma\equiv \Gamma[\phi,\Delta;\emptyset]$ to be understood as the 2PI effective action to avoid complicating the notation.

If we assume that the configuration space is spanned by the coordinates $\{J_a, K_\alpha\}$ and, alternatively, that we can describe the same configuration space by the classical fields $\{\phi^a, \Delta^\alpha\}$, we can use the chain rule to derive 
\begin{equation}
    \begin{split}
        \delta^a_{\ b} = \fderiv{\phi^a}{\phi^b} =& \, \fderiv{\phi^a}{J_c} \fderiv{J_c}{\phi^b} + \fderiv{\phi^a}{K_{\gamma}} \fderiv{K_{\gamma}}{\phi^b} \\
        =& \, \fderiv{^2 \mathcal{W}}{J_a \delta J_c} \fderiv{^2 \Gamma}{\phi^c \delta \phi^b} + \fderiv{^2\mathcal{W}}{J_a \delta K_{\gamma}} \fderiv{^2\Gamma}{\Delta^{\gamma} \delta \phi^b} \, .
    \end{split}
    \label{eqn:2piid}
\end{equation}
Without the latter term, this is the standard relation used in the derivation of the Wetterich equation. Here, we need to combine this relation with other identities derived from the remaining variables to close the equations. 

With the notation introduced above, the convexity conditions of the 2PI effective action can be concisely expressed as
\begin{equation}
    \delta^A_{\ B} = \fderiv{\mathcal{P}^A}{\mathcal{P}^B} = \fderiv{\mathcal{P}^A}{\mathcal{Q}_C} \fderiv{\mathcal{Q}_C}{\mathcal{P}^B} = \fderiv{^2 \mathcal{W}}{\mathcal{Q}_A \delta \mathcal{Q}_C} \fderiv{^2 \Gamma}{\mathcal{P}^C \delta \mathcal{P}^B} \, .
    \label{eqn:2PImasterconvex}
\end{equation}
Distinguishing one- and two-point objects, \cref{eqn:2PImasterconvex} becomes an operator-valued $2 \times 2$ matrix. From its four components, we identify the four identities (first appearing in endnote 11 of~\cite{Cornwall:1974vz}, see also~\cite{Millington:2021ftp})
\begin{equation}
    \begin{split}
        \delta^a_{\ b} =& \, \fderiv{^2 \mathcal{W}}{J_a \delta J_c} \fderiv{^2 \Gamma}{\phi^c \delta \phi^b} + \fderiv{^2 \mathcal{W}}{J_a \delta K_\gamma} \fderiv{^2 \Gamma}{\Delta^\gamma \delta \phi^b} \, , \\
        \delta^\alpha_{\ \beta} =& \, \fderiv{^2 \mathcal{W}}{K_\alpha \delta J_c} \fderiv{^2 \Gamma}{\phi^c \delta \Delta^\beta} + \fderiv{^2 \mathcal{W}}{K_\alpha \delta K_\gamma} \fderiv{^2 \Gamma}{\Delta^\gamma \delta \Delta^\beta} \, , \\
        0 =& \, \fderiv{^2 \mathcal{W}}{K_\alpha \delta J_c} \fderiv{^2 \Gamma}{\phi^c \delta \phi^b} + \fderiv{^2 \mathcal{W}}{K_\alpha \delta K_\gamma} \fderiv{^2 \Gamma}{\Delta^\gamma \delta \phi^b} \, , \\
        0 =& \, \fderiv{^2 \mathcal{W}}{J_a \delta J_c} \fderiv{^2 \Gamma}{\phi^c \delta \Delta^\beta} + \fderiv{^2 \mathcal{W}}{J_a \delta K_\gamma} \fderiv{^2 \Gamma}{\Delta^\gamma \delta \Delta^\beta} \, .
    \end{split}
    \label{eqn:2PIconvexity}
\end{equation}
The first equation of \cref{eqn:2PIconvexity} is \cref{eqn:2piid}.
However, we might prefer working directly with \cref{eqn:2PImasterconvex}. 

To find an expression for $\Delta^\alpha$ and close the flow, we note that the relation between the two-point function and the Schwinger functional is
\begin{equation}
    \fderiv{^2\mathcal{W}}{J_a \delta J_b} = \Delta^{a b} - \phi^a \phi^b \, .
\end{equation}
We can obtain an expression only involving $\Gamma$ by noting that, according to \cref{eqn:2PImasterconvex},
\begin{equation}
    \fderiv{^2 \mathcal{W}}{\mathcal{Q}_A \delta \mathcal{Q}_B} = g^{A B} \, ,
    \label{eqn:prop2}
\end{equation}
with
\begin{equation}
    g_{A B} = \fderiv{^2 \Gamma}{\mathcal{P}^A \delta \mathcal{P}^B} \, .
    \label{eqn:prop}
\end{equation}
This implies,
\begin{equation}
    \Delta^{a b} - \phi^a \phi^b = g^{a b} \, .
    \label{eqn:Delta}
\end{equation}
An explicit expression can be obtained by splitting the capital Latin indices into one and two-point indices. We then write \cref{eqn:prop2} as the inverse of \cref{eqn:prop}. This boils down to the inversion of an operator valued $2 \times 2$ matrix. Inserting this into \cref{eqn:2PImasterconvex}, we find
\begin{equation}
    g^{a b} \left[ \fderiv{^2 \Gamma}{\phi^b \delta \phi^c} - \fderiv{^2 \Gamma}{\phi^b \delta \Delta^{\delta}} \left( \xi^{-1} \right)^{\delta \epsilon} \fderiv{^2 \Gamma}{\Delta^{\epsilon} \delta \phi^c} \right] = \delta^a_{\ c} \, ,
\end{equation}
\begin{equation}
    \xi_{\alpha\beta} = \fderiv{^2 \Gamma}{\Delta^{\alpha} \delta \Delta^{\beta}} \, .
\end{equation}
By using this additional identity, we can indeed close the flow equation \cref{eqn:2PIflow}.

Taking a $t$-derivative of \cref{eqn:Delta} and using the chain rule, we can derive the following expression involving the $t$-derivative of the two-point function:
\begin{equation}
    \left[ \bm{1}^{a b}_{\ \ c d} - \fderiv{g^{a b}}{\Delta^{c d}} \right] \partial_t \Delta^{c d} = 0 \, .
    \label{eqn:gauget}
\end{equation}
Thus, the $t$-derivative of the two-point function must be an eigenvector of the operator in \cref{eqn:gauget} with vanishing eigenvalue. Note that this does not fix the $t$-derivative of the two-point function, and this is related to the fact that we can choose different regulators implementing different RG derivatives for the two-point function.

%%%%%%%%%%%%%%%%%%%%

\section{Commutation of Derivatives}
\label{sec:derivs}

When working with $n$PI effective actions and considering derivatives which keep different objects constant, care should be taken when commuting partial derivatives. For example, we can work in the natural variables $\phi^a$ and $\Delta^\alpha$ of the 2PI effective action. Derivatives with respect to those commute with each other,
\begin{equation}
    \begin{split}
        \left( \fderiv{}{\phi^a} \right)_{\Delta^\alpha} \left( \fderiv{}{\Delta^\alpha} \right)_{\phi^a} =& \, \left( \fderiv{}{\Delta^\alpha} \right)_{\phi^a} \left( \fderiv{}{\phi^a} \right)_{\Delta^\alpha} \\
        \equiv&\, \fderiv{^2}{\phi^a \delta \Delta^\alpha} \, ,
    \end{split}
    \label{eqn:2ndderivnat}
\end{equation}
where we use brackets to denote explicitly which variables are kept constant.

A less trivial example is encountered when including a derivative by $K_\alpha$ that keeps $J_a$ constant. Note that this derivative encodes the RG derivative in the $2$PI effective action. The commutator of it with $\smash{\fderiv{}{\phi^a}}$ arises, e.g. when taking $\phi^a$ derivatives of the flow equation \cref{eqn:2PIflow}. It can be shown that the commutator of both only vanishes if
\begin{equation}
    \left( \fderiv{}{\phi^a} \right)_{\Delta^\alpha} \fderiv{\Delta^\beta}{K_\alpha}\stackrel{?}{=} 0 \, .
    \label{eqn:derivcond}
\end{equation}
In general, this is not the case. For example, let us consider a zero-dimensional case with $\Delta (\phi, K) = \phi K^2$. Since a $\phi$ derivative of $\Delta$ with $\Delta$ kept fixed is zero, we have
\begin{equation}
    \left( \pderiv{}{K} \right)_{\phi} \left( \pderiv{\Delta}{\phi} \right)_\Delta = 0 \, .
\end{equation}
However, interchanging the derivatives yields
\begin{equation}
    \left( \pderiv{}{\phi} \right)_\Delta \left( \pderiv{\Delta}{K} \right)_{\phi} = K \neq 0 \, .
\end{equation}
Thus, \cref{eqn:derivcond} is not fulfilled and the derivatives do not commute, which we have shown explicitly. More generally, we expect both derivatives to commute only for special relationships between $\Delta$, $\phi$, and $K$. To make this point clearer, we translate the condition \cref{eqn:derivcond} to geometrical objects. Using
\begin{equation}
    K_\alpha = \fderiv{\Gamma}{\Delta^\alpha} \, ,
\end{equation}
we can derive
\begin{equation}
    \left( \fderiv{\Delta^\alpha}{K_\beta} \right)_\phi = \left( \fderiv{^2 \Gamma}{\Delta^\alpha \delta \Delta^\beta} \right)^{-1} \, .
\end{equation}
Inserting this in \cref{eqn:derivcond}, we find
\begin{equation}
    \left( \fderiv{^2 \Gamma}{\Delta^\alpha \delta \Delta^\beta} \right)^{-1} \fderiv{^3 \Gamma}{\Delta^\beta \delta \Delta^\gamma \delta \phi^a} \left( \fderiv{^2 \Gamma}{\Delta^\gamma \delta \Delta^\delta} \right)^{-1} \stackrel{?}{=} 0 \, .
\end{equation}
Assuming the invertibility of the metric in the $\Delta$-$\Delta$ sector, \cref{eqn:derivcond} is equivalent to the vanishing of a part of the Amari--Chentsov tensor, i.e., 
\begin{equation}
    \hat{\varGamma}_{a \alpha \beta} = 0 \, .
    \label{eqn:derivcondgeo}
\end{equation}
While this identity is fulfilled for free QFTs, generic QFTs will violate \cref{eqn:derivcondgeo} and lead to a non-commutativity of derivatives through \cref{eqn:derivcond}. This subtlety must be taken into account when applying $\phi^a$ derivatives on \cref{eqn:2PIflow}.

%%%%%%%%%%%%%%%%%%%%

\bibliography{bib}

\end{document}